\documentclass[journal]{IEEEtran}
\IEEEoverridecommandlockouts
\usepackage{times,amsmath,color,amssymb,graphicx,epsfig,cite,psfrag,subfigure,algorithm,balance}
\usepackage{amsfonts,pifont,enumerate,cases}
\usepackage{mathrsfs} 
\usepackage[table]{xcolor} 
\usepackage{verbatim} 
\usepackage{bm}
\usepackage{cuted,stfloats}
\usepackage{algorithm}
\usepackage{algorithmic}

\usepackage{longtable}
\usepackage{blindtext}
\usepackage{multirow}
\usepackage{float}
\usepackage{threeparttable}
\usepackage{makecell}
\usepackage[utf8]{inputenc}
\usepackage{url}
\usepackage{booktabs}
\usepackage{amssymb}
\usepackage{bbding}
\usepackage{pifont}
\usepackage{wasysym}
\usepackage{utfsym}
\usepackage{fontawesome}
\usepackage[algo2e,ruled,vlined,linesnumbered,lined,boxed,commentsnumbered]{algorithm2e}
\usepackage{amsmath,mathtools}
\usepackage[
    colorlinks=true,
    linkcolor=blue,
    citecolor=blue,
    urlcolor=magenta
]{hyperref}
\usepackage{array}

\begin{document}
\title{Movable Antennas Enabled Wireless Powered Networks: Principles and Technologies}
\author{Zhendong Li, Yiran Zheng, Tianyu Li, Zhou Su, Wen Chen, and Ying Wang
\thanks{Zhendong Li, Yiran Zheng and Tianyu Li are with the School of Information and Communication Engineering, Xi'an Jiaotong University, Xi'an 710049, China (email: lizhendong@xjtu.edu.cn; 18205322292@stu.xjtu.edu.cn; buptlty@stu.xjtu.edu.cn). Zhou Su is with the School of Cyber Science and Engineering, Xi'an Jiaotong University, Xi'an 710049, China (email: zhousu@ieee.org). Wen Chen is with the Department of Electronic Engineering, Shanghai Jiao Tong University, Shanghai 200240, China (e-mail: wenchen@sjtu.edu.cn). Ying Wang is with the State Key Laboratory of Networking and Switching Technology, Beijing University of Posts and Telecommunications, Beijing 100876, China (e-mail: wangying@bupt.edu.cn). (Corresponding author: Zhou Su)}
\vspace{-1.5em}
}
\maketitle
\thispagestyle{empty}

		\maketitle
		
\begin{abstract}
As an emerging framework, movable antenna (MA)-enabled wireless powered networks (WPNs) have attracted growing attention. WPNs integrate wireless communication and energy transfer. MA can dynamically adjust the position of antenna units by introducing additional spatial degrees of freedom, so as to make full use of channel gain, optimize the effect of energy beamforming, and further improve the performance of WPNs. In this article, we first classify the implementations of MA by driving architectures, and review the fundamental principles of WPNs. We then highlight the key advantages of MA-enabled WPNs in enhancing wireless power transfer efficiency, realizing flexible and adaptive beamforming, and improving system robustness and interference resilience. Furthermore, four representative application scenarios and three key enabling technologies are discussed. A case study is also presented to show the reduction in transmit power brought by MA for WPNs. Finally, we discuss the challenges and future directions of MA-enabled WPNs, aiming to provide reference for future research and practice.

\end{abstract}
		
			
		
\section{Introduction}
 
\IEEEPARstart{A}{s} the new generation of mobile communication technology, sixth-generation (6G) networks are envisioned to realize the Internet-of-everything (IoE), connecting a massive number of heterogeneous devices, sensors, and intelligent terminals. Compared with fifth-generation (5G) networks, 6G is expected to support a substantially larger number of connected devices, driven by the rapid growth of the Internet-of-things (IoT) and emerging data-intensive applications\cite{Shen_2022_6GHRM}. In this context, wireless powered networks (WPNs) show significant technical value. Addressing the common pain points of massive devices in 6G scenarios, such as limited battery life, high replacement costs, and restricted access to reliable energy sources, WPNs can achieve remote wireless power supply through wireless power transfer (WPT) technology, enabling these devices to break away from their dependence on traditional batteries and achieve sustainable operation. More importantly, WPNs enable the integrated delivery of wireless information and energy, offering a promising paradigm for sustaining large-scale and energy-constrained IoT deployments in future 6G networks.

Despite the great potential of WPNs, their practical performance remains constrained by conventional antenna architectures. In particular, existing WPNs typically rely on fixed-position antennas (FPAs) with predetermined locations, orientations, and radiation patterns, which provide limited spatial degrees of freedom (DoF) and restrict coverage flexibility. In the face of dynamic mobile terminals or complex multipath fading, the conventional FPAs are difficult to accurately align with the energy receiver, which leads to large loss of radio frequency (RF) energy path and low WPT efficiency\cite{Yan_2025_PIMRC_ISAC}. Moreover, the FPAs are difficult to fully cope with the uneven distribution of terminals and obstacles, resulting in a significant decline in the stability and reliability of WPNs' energy supply. They are also unable to achieve continuous wireless energy supply services. To sum up, the inherent spatial rigidity of FPA-based architectures fundamentally limits the efficiency, flexibility, and adaptability of WPNs, motivating the development of new antenna technologies that can provide additional spatial DoF.

To address these limitations, movable antenna (MA), as an innovative technical solution, has been introduced into WPNs. MA can dynamically adjust the antenna unit through mechanical or electronic control, so that the antennas can match the channel environment and terminal position in real time, so as to break the spatial constraints of FPAs\cite{Zhu_2023_MovableAnt}. This spatial flexibility allows the antenna configuration to adapt to the locations of energy receivers and the surrounding channel conditions. In particular, MA can improve the spatial alignment between energy transmitters and receivers, thereby mitigating the adverse effects of path loss and multipath fading, and enhancing energy harvesting efficiency. Meanwhile, MA can effectively expand the spatial DoF of the antennas, make full use of the channel gain, and optimize the effect of energy beamforming. This means that WPNs can use more flexible beamforming technology to accurately focus energy in a specific direction, significantly improving energy intensity. In addition, MA can reconstruct antenna deployment and radiation pattern in real time according to terminal movement, obstacle occlusion and network load changes to adapt to the dynamic changes of the environment, so as to improve the energy supply stability and interference resilience of WPNs.

    Although MA-enabled WPNs have significant technical advantages and promising development prospects, research on MA-enabled WPNs is still at the initial stage. Nevertheless, existing research suffers from several common limitations. Most existing works adopt idealized assumptions, such as zero positioning delay and negligible mechanical power consumption for MA movement, which may lead to over‑optimistic performance predictions. In addition, prior reviews mainly focus on general MA communication systems, while few provide dedicated summaries for WPN use cases.  \cite{Zhao_2026_ReconfigAntenna} provided a unified taxonomy of various reconfigurable antennas, including MAs and fluid antennas, with an emphasis on cross-paradigm comparisons. \cite{Bepari_2026_FluidAntenna} presented a comprehensive cross-layer survey of fluid antenna systems, with MAs mainly serving as a comparative reference. While these studies offer valuable insights into reconfigurable antenna technologies, a systematic and dedicated review of MA-enabled WPNs, covering their system architectures, key enabling techniques, and use cases, remains relatively unexplored.
    This article focuses on MA-enabled WPNs, and systematically expounds its core working principle, opportunities and advantages, typical application scenarios, key supporting technologies, challenges and future research directions. Specifically, Section \ref{sec:2} introduces the architectures and implementations of MA, followed by an overview of two representative WPN paradigms, namely simultaneous wireless information and power transfer (SWIPT) and wireless powered communication network (WPCN). The key advantages of MA-enabled WPNs are then highlighted in Section \ref{sec:3}, while their typical application scenarios are discussed in Section \ref{sec:4}. In Section \ref{tec}, three key technologies used in MA-enabled WPNs are analyzed. Section \ref{sec:5} presents a case study on transmit power minimization in MA-assisted SWIPT networks, where a tractable alternating optimization (AO) framework is developed to demonstrate the performance gains enabled by MA. Finally, the challenges and potential research directions are discussed in Section \ref{sec:6}, and the article is concluded in Section \ref{sec:7}.

\section{Principles of MA and Wireless Powered Networks}
\label{sec:2}

\subsection{MA Architectures and Implementations}

In this subsection, we mainly introduce the architectures and different implementations of MA. MA is an emerging wireless communication hardware technology. Its core advantage is that it can dynamically adjust the physical position of the antenna unit, so as to actively optimize the wireless channel conditions to adapt to the changing communication environment. This flexibility enables MA to show great potential in reducing path loss and interference, enhancing beamforming gains, and improving the overall performance of wireless communication and WPT systems. Different from conventional FPAs, MA significantly improves the performance of wireless communication system by introducing additional spatial DoF\cite{Li_2025_MovableAntennasISAC}.

The practical realization of MA depends on advanced precision driving and micro-machining technologies, and its basic types are mainly classified according to the driving mechanism and the dimensions of reconfigurability. Currently, the primary implementation mechanisms include micro electro mechanical system (MEMS), stepper motor drive, and fluid drive. Table \ref{tab:comparison} presents a performance comparison of the three different driving mechanisms. MEMS uses semiconductor micromachining technology to achieve small and accurate displacement or shape changes of antenna elements, which is suitable for scenes extremely sensitive to size and power consumption\cite{Brown_1998_RFMEMS}. MA driven by stepper motor adopts the traditional electromechanical servo system, which can drive the antenna unit for macro and large-scale discrete position adjustment. It is suitable for scenes that require a large moving range but are not sensitive to size and weight\cite{Wu_2025_MAOptimal}. As a new biological heuristic method, fluid driven technology can use fluid pressure to drive flexible or deformable structures to change the shape or position of the antenna unit. It has great application potential in the field of wearable or implantable medical devices, as well as in the aerospace industry. 

Beyond MA, other representative reconfigurable antenna paradigms have also been developed for next generation wireless systems\cite{Zhao_2026_ReconfigAntenna}. Fluid antennas adjust radiation ports via liquid‑metal or pixel based reconfiguration, achieving spatial diversity without large scale physical displacement. Pinching antennas realize reconfigurable radiation points by inserting or squeezing dielectric units along dielectric waveguides, and are particularly suited for millimeter wave line‑of‑sight dominated scenarios. Different from the mechanically driven reconfiguration of MA, reconfigurable holographic antennas adopt solid state metasurface tuning to reshape aperture field patterns, which enables fast beam manipulation without macroscopic mechanical motion. Overall, these implementation architectures provide different tradeoffs among positioning accuracy, moving range, response speed, hardware complexity, and power consumption. Their respective characteristics make them suitable for different application environments, and all such spatially reconfigurable designs, including MA, possess the spatial flexibility that distinguishes MA from conventional FPAs. Such hardware-level spatial flexibility provides the foundation for exploiting the additional spatial DoF of MA in WPNs, where antenna positioning can be jointly optimized with energy beamforming to improve WPT efficiency.

\begin{table}[ht]
\caption{Comparison of Different Driving Mechanisms for MA}
\label{tab:comparison}
\centering
\begin{tabular}{m{0.195\linewidth} m{0.236\linewidth} m{0.18\linewidth} m{0.2\linewidth}}
\hline
\textbf{Characteristic} & \textbf{MEMS} & \textbf{Stepper Motor Drive} & \textbf{Fluid Drive} \\
\hline
\textbf{Moving Range} & Micron Level & Centimeter to Meter Level & Millimeter to Centimeter Level \\
\hline
\textbf{Driving Force} & Weak & Strong & Medium \\
\hline
\textbf{Response Speed} & Fast & Slow & Medium \\
\hline
\textbf{Power Consumption} & Low & High & Medium \\
\hline
\textbf{Integration} & High & Low & Medium \\
\hline
\end{tabular}
\end{table}

 \noindent Note: The qualitative ratings in this table are indicative and may differ for practical hardware implementations.

\subsection{Primer on Wireless Powered Networks}

WPNs refer to network systems that provide remote energy supply to electronic devices through wireless RF signals to support their continuous operation, which have two key implementation technologies: SWIPT and WPCN. SWIPT and WPCN can be applied in fields such as IoT, mobile communication, and special environment communication, providing a crucial theoretical basis and technical path for the realization of IoE\cite{Xu_2023_RIS_SWIPT}.

\begin{figure}[t]
    \centering
   \includegraphics[width=0.49\textwidth]{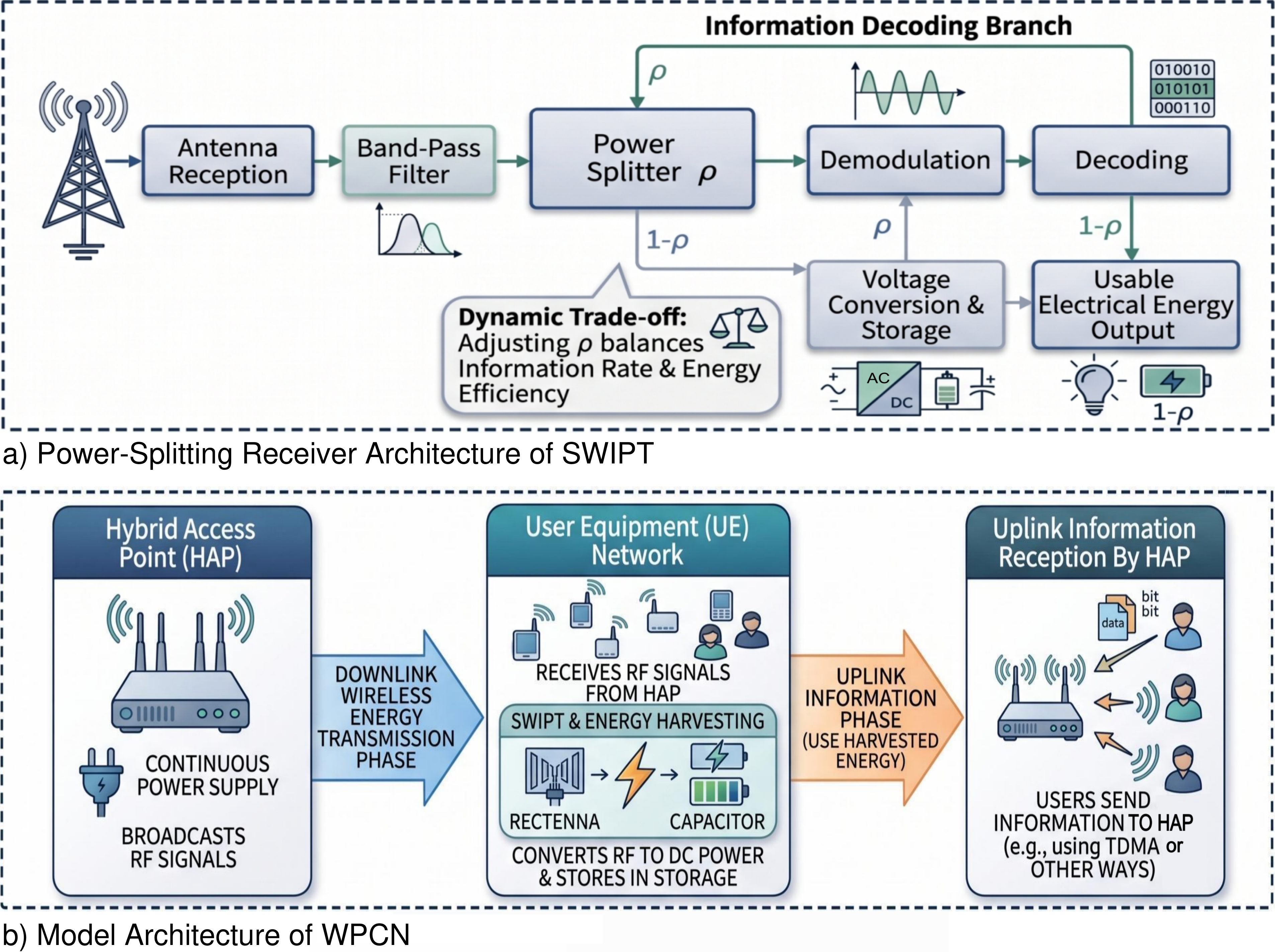} 
    \caption{Model architecture of SWIPT and WPCN. Subfigure (a) shows the receiver-level SWIPT power-splitting architecture, and (b) presents the network-level transmission workflow of WPCN.}
    \label{fig:workflow}
\end{figure}

SWIPT allows receiving devices to simultaneously acquire information and energy from the same RF signal\cite{Zhang_2013_SWIPT}. It breaks the boundary between information transmission and energy supply in traditional communication, and separates the received signals into information flow for communication and energy flow for charging at the same time or in sequence, so as to maximize the utilization efficiency of spectrum and hardware. Taking the power-splitting architecture of SWIPT receiver as an example, the specific working mode is shown in Fig. \ref{fig:workflow}(a). The received signal is dynamically divided into two paths according to the proportion $\rho$  and $1-\rho$ through a power splitter, where $\rho$ denotes the power splitting ratio. One path is sent to the information decoder, and then restored to digital information through demodulation, decoding and other modules in turn. The other is fed into the energy harvesting circuit, where the received RF signal is converted into direct current (DC) through rectification and subsequent voltage conversion. The harvested energy can then be directly utilized or stored in a battery or capacitor.

WPCN, in contrast, constructs a network-level framework based on WPT in which wireless energy harvesting and information transmission are coordinated over different time periods\cite{Wang_2016_TwoWayRelay}. Fig. \ref{fig:workflow}(b) shows the system model implementation of WPCN. The core component of the system is the hybrid access point (HAP), which is usually equipped with a continuous power supply and multiple antennas, and is responsible for downlink WPT and uplink information reception. The HAP first broadcasts energy to all users in the network through RF signals, that is the downlink WPT phase. Each user then harvests the received RF energy and stores the resulting electrical energy in a battery or capacitor. During the subsequent uplink information transmission phase, the users utilize the harvested energy to transmit their information to the HAP through time-division multiple access or other multiple-access schemes. Through the coordination of WPT and information transmission, WPCN provides a practical framework for enabling energy self-sufficiency in wireless networks.

\section{Opportunities and Advantages of MA-Enabled WPNs}
\label{sec:3}
In this section, we mainly analyze the advantages of MA-enabled WPNs, which include: enhanced WPT efficiency, flexible and adaptive beamforming, and improved robustness and interference resilience.

\subsection{Enhanced WPT Efficiency}

By dynamically adjusting the positions of antenna elements, MA can exploit the spatial variations of wireless channels and thereby improve the WPT efficiency of WPNs. In traditional WPNs, HAP is usually equipped with FPAs, whose energy beamforming capability is constrained by a predetermined antenna configuration. This spatial rigidity becomes particularly problematic in dynamic wireless environments. Specifically, on the one hand, wireless channels are highly susceptible to environmental interference, leading to issues such as multipath fading, blockage, and shadowing. On the other hand, terminal devices in the network have mobile characteristics and their locations are not fixed. FPAs are difficult to adapt to dynamically changing channel environments and are unable to consistently provide stable and efficient energy supply for mobile devices.

However, the introduction of MA breaks this restriction. Specifically, when the channel quality deteriorates because of fading or blockage, antenna elements can be repositioned toward locations with more favorable channel conditions. By exploiting the spatial variations of the channel over the antenna moving region, MA can avoid unfavorable channel conditions, mitigate energy losses, and establish more efficient WPT links. At the same time, the dynamic position optimization method can optimize the beam transmission path, strengthen the focusing ability of the energy beam, and build a low-loss and high-stability WPT link. In addition, the position adjustable antenna structure can adapt to the position change of the mobile terminal. It can match the receiving position of devices in real time, optimize the WPT path, and further improve the energy capture efficiency. To sum up, MA optimizes WPNs from the spatial dimension by introducing additional spatial DoF. Through adaptive antenna positioning and beamforming, MA-enabled WPNs can improve WPT efficiency while providing greater adaptability to channel variations, blockage, and user mobility.

\subsection{Flexible and Adaptive Beamforming}

Conventional FPA-based systems have limited spatial flexibility, making it challenging to adapt energy and communication beams to dynamically changing user locations and channel conditions. In contrast, the mobility of antenna elements in MA introduces additional spatial DoF for beamforming optimization. By jointly adjusting antenna positions and beamforming weights, MA-enabled WPNs can more flexibly control the spatial distribution and directionality of transmitted signals, thereby improving both wireless communication and power transfer performance. 

In practical WPNs, channel conditions can vary rapidly because of user mobility, multipath propagation, and dynamic blockage. A fixed antenna configuration may therefore become mismatched with the desired beam direction as users move. MA can actively adjust the relative spatial position of each antenna unit and dynamically optimize the radiation pattern of the antenna array, so as to obtain higher beamforming gain in the desired direction and strengthen the reception strength of effective signals. Moreover, it can also actively suppress the signal radiation in the direction of non-target node and interference source, and weaken the signal leakage in the direction of non-target. This advantage is particularly obvious in the network scenario of multi-user coexistence. It can effectively reduce the signal interference between different terminals, improve the signal-to-noise ratio of the system, and ensure the stability and reliability of the communication link. For WPT, the additional spatial DoF provided by MA enable more precise and adaptive energy beamforming. By jointly optimizing antenna positions and energy beamforming, RF energy can be concentrated more effectively toward intended energy receivers, thereby reducing energy leakage and spatial diffusion. Even if there is occlusion, channel fluctuation or device displacement in the environment, the system can still provide stable energy supply for distributed terminals and mobile devices. To sum up, MA breaks through the fixed limitation of traditional beamforming. With its flexible spatial adjustment ability, it realizes adaptive beam optimization, and improves the quality of wireless communication and the stability of WPT.

\subsection{Improved Robustness and Interference Resilience}

For WPNs with FPAs, after the antenna deployment is completed, the physical positions of antenna units cannot be changed. Also, the channel transmission conditions remain basically unchanged. This rigid structure makes the system prone to signal fading, obstacle shielding and other problems in the complex and changeable electromagnetic environment. The system is unable to make dynamic adjustment in time, resulting in its poor transmission stability. In addition, FPAs lack spatial adjustability and flexibility, so they can hardly mitigate external electromagnetic interference, and their interference resilience has obvious shortcomings. 

The introduction of MA effectively solves the above pain points and greatly optimizes the comprehensive performance of WPNs. Specifically, MA can rely on the unique spatial DoF to complete dynamic optimization, move flexibly in limited space, and make full use of the spatial changes of wireless channels to improve the stability of system operation. During operation, antenna units can adjust their physical positions in real time, and continuously optimize the signal transmission link. It can actively avoid the transmission blind zone and deep fading area, and weaken the negative impact of adverse channel conditions. This can ensure smooth signal transmission and maintain long-term and stable wireless communication and WPT. Meanwhile, MA also features excellent interference resilience. The system can effectively weaken the adverse effects of external interference signals by fine-tuning the spatial positions of the antennas and changing the signal radiation and reception paths. To sum up, MA enhances the robustness and interference resilience of WPNs by enabling adaptive spatial reconfiguration, making them more suitable for complex and dynamically changing wireless environments.

\section{Application Scenarios of MA-Enabled WPNs}
\label{sec:4}

\begin{figure}[htbp]
    \centering
   \includegraphics[width=0.49\textwidth]{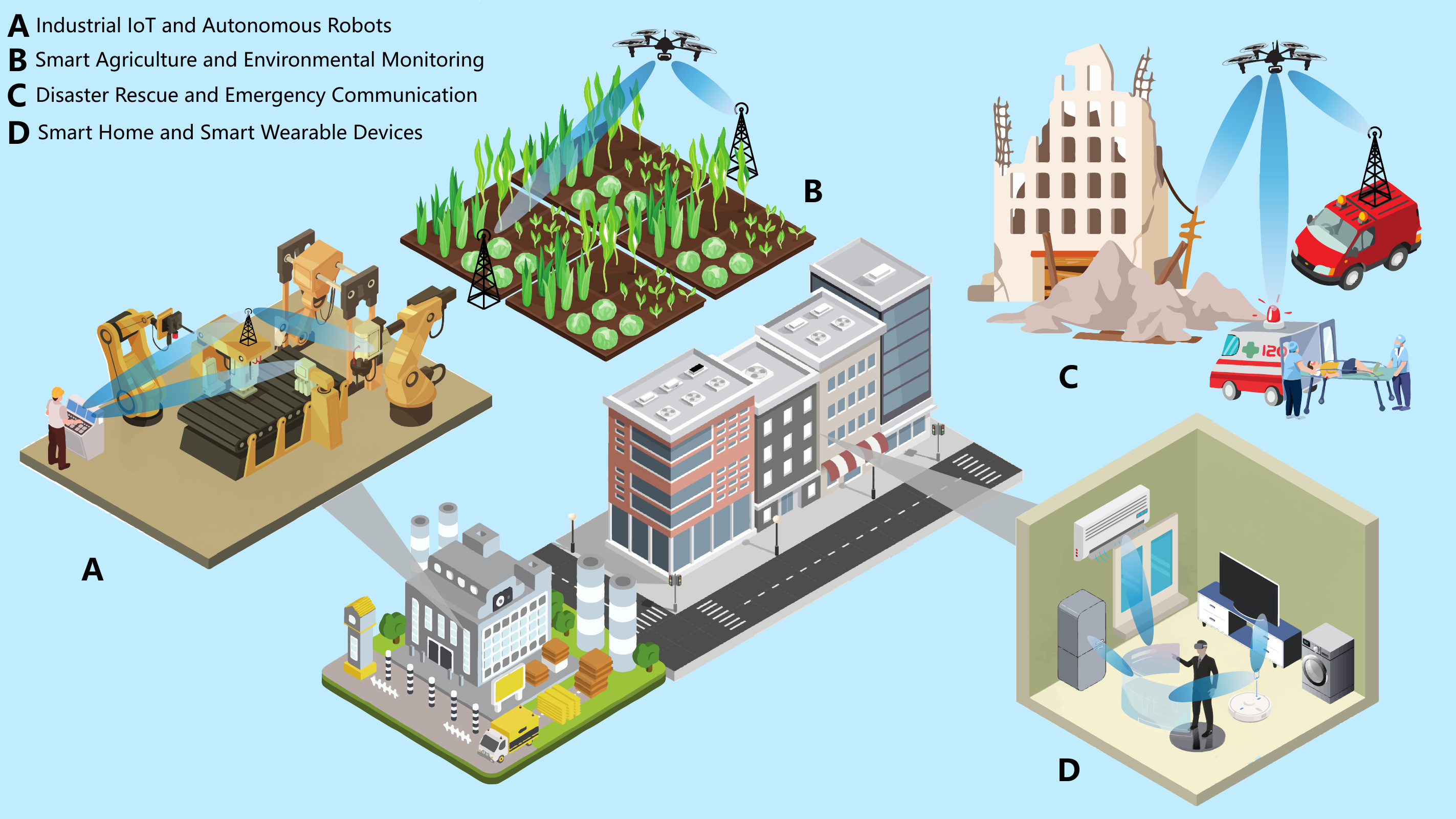} 
    \caption{Application scenarios of MA-enabled WPNs.}
    \label{fig:applications}
\end{figure}

The unique spatial flexibility of MA makes them well suited to a wide range of wireless powered networking scenarios, particularly those involving mobile users, dynamic environments, and stringent energy requirements. As illustrated in Fig. \ref{fig:applications}, representative application scenarios of MA-enabled WPNs include the following.

\subsection{Industrial IoT and Autonomous Robots}

In smart factories, MA-based HAPs or relay nodes can be deployed on ceilings, production-line equipment, or robotic platforms to provide dynamic and continuous wireless energy supply for a large number of industrial devices. Inspection robots, automated guided vehicles, and wireless sensors often operate under strict mobility and energy constraints, while wired charging restricts their mobility and conventional wireless charging suffers from limited coverage and charging efficiency. By deploying energy beacons with MA, energy transmitters can adapt their antenna configurations to the locations of mobile robots and sensor nodes, enabling more targeted wireless power delivery. Moreover, in the face of the interference and electromagnetic reflection of a large number of metal equipment in factories, MA array can adjust the beam direction and shape in real time, and optimize the beamforming, thereby mitigating the effects of blockage and fading. This capability can improve the efficiency and reliability of WPT while supporting the continuous operation of mobile industrial devices\cite{Rosabal_2026_MA_WET_IoT}. 

\subsection{Smart Agriculture and Environmental Monitoring}

In smart agriculture and environmental monitoring, wireless sensors are often distributed over large and remote areas, where frequent battery replacement or wired power supply is costly and impractical. MA-enabled mobile energy transmitters, such as unmanned aerial vehicles (UAVs) and ground vehicles, can be deployed in farmland or nature reserves to regularly supply wireless energy for widely distributed soil sensors, meteorological monitoring points and pest monitoring points. MA can flexibly adjust the direction and intensity of the energy beam according to the geographical distribution and energy demand of the sensors to achieve regional or point-to-point WPT. It can also dynamically find favorable transmission path according to crop growth, or reconstruct beamforming by adjusting the hovering position of UAVs equipped with MA array. Thus it can always maintain high-quality connection with monitoring points, maximize WPT efficiency, and adapt to environmental changes. This not only reduces the dependence of sensors on batteries and the maintenance costs, but also ensures the long-term stable operation of sensor networks, providing reliable data support for precision agriculture management and environmental early warning\cite{Chien_2021_UAV_WPT_AG}.

\subsection{Disaster Rescue and Emergency Communication}

Natural disasters such as earthquakes and floods can severely damage conventional power and communication infrastructure, leaving rescue personnel and emergency communication devices without reliable energy supplies. In such extreme situations, MA-enabled emergency communication vehicles, tethered UAVs, or portable base stations can be rapidly deployed to the disaster area to establish an emergency communication network and provide wireless energy for handheld devices used by rescue personnel. MA-based energy transmitters can dynamically reconfigure their antenna positions and beamforming patterns to identify favorable transmission conditions and maintain reliable energy links. UAVs equipped with MA array can provide continuous energy for emergency communication equipment on the ground by optimizing their trajectory and beamforming, while ensuring the safety of communication\cite{Liu_2025_UAVMovableAntennaArray}. This capability enables rapid, on-demand, and targeted WPT, thereby extending the operating time of emergency equipment and improving the efficiency of disaster response.

\subsection{Smart Home and Smart Wearable Devices}

 In a smart home environment, smart devices can be embedded with micro energy receivers. By integrating MA-enabled energy transmitters into ceilings, walls, or other indoor infrastructure, the system can monitor the positions of the devices in the room. Then it can automatically adjust the antenna beams, and transmit the energy to the smart devices requiring charging, without any operation by the user. MA can dynamically adjust the optimal positions and directions of antennas based on the locations of receiving devices and channel states to align the energy beam precisely with receiving terminals, and suppress energy leakage as well as sidelobe interference. This approach can overcome the obstruction of obstacles and reduce unnecessary radiation to human body\cite{Murata_2026_HumanAwareBeam}. In future indoor deployments of MA‑enabled WPNs, such non‑intrusive wireless charging can prolong device battery lifetime and improve the practical usability of smart home systems.

\section{Key Technologies}
\label{tec}

Compared to traditional WPNs, MA breaks through the static channel constraint and WPT efficiency bottleneck through dynamic scheduling of antenna spatial DoF, effectively enhancing network energy supply stability and energy utilization efficiency. Its core performance gain is mainly achieved by three types of key technologies, including MA position optimization, energy beamforming and spatial energy focusing, and position-dependent channel estimation and prediction. Specifically, accurate position-dependent channel estimation provides indispensable channel state information, which serves as the fundamental input to implement MA position optimization and energy beamforming design. In this section, we mainly elaborate on the above core technologies.

\subsection{MA Position Optimization}

MA position optimization is a fundamental technology for realizing the spatial flexibility of MA-enabled WPNs. Unlike the fixed antenna configuration of conventional WPNs, MA can realize the precise spatial matching of WPT link through the dynamic position fine-tuning of antenna arrays in the limited space. In conventional WPNs, the fixed antenna layout is easy to cause problems such as uneven cell energy coverage, large fluctuations in equipment receiving energy gain, and serious energy attenuation of edge nodes. By flexibly adjusting the spatial position of the transmitting antennas, MA can effectively avoid the zero point of channel fading and optimize the channel power gain between base stations and different energy acquisition devices. In multi-user WPNs, MA position optimization can further balance overall WPT efficiency and energy fairness among users while maintaining reliable energy supply. It combines hardware constraints such as antenna movement range and minimum element spacing to determine the optimal antenna position configuration. By dynamically adapting the distribution of terminals to compensate for the spatial attenuation and fading loss of WPT, the energy coverage and effective energy supply efficiency of WPNs can be significantly improved without additional transmit power.

\subsection{Energy Beamforming and Spatial Energy Focusing}

Energy beamforming and spatial energy focusing are key signal processing technologies for delivering RF energy efficiently and directionally in MA-enabled WPNs. The core objective of these technologies is to realize the spatial aggregation of wireless radiation energy, suppress energy leakage and invalid radiation, and enhance the energy harvesting efficiency of terminals. Conventional FPA-based WPNs are constrained by fixed array geometries, which limit their ability to adapt beam coverage and may result in energy leakage and inefficient spatial utilization. Relying on the advantage of dynamically adjustable spatial DoF of antenna positions, MA can realize the joint optimization of array geometry and beam weight. By accurately adjusting the beam direction, beam width and power distribution, MA can highly focus electromagnetic energy on the target terminal area, forming a directional and energy-efficient WPT channel. For multi-user scenarios, such joint optimization can also reduce inter-user energy coupling and adapt the spatial focusing region to heterogeneous energy demands. Consequently, MA-enabled energy beamforming provides more targeted and efficient WPT for distributed and mobile devices.

\subsection{Position-Dependent Channel Estimation and Prediction}

Position-dependent channel estimation and prediction provide essential channel knowledge for MA position optimization and energy beamforming. Continuous fine-tuning of the antenna positions will directly change the channel gain, spatial correlation and link attenuation characteristics of WPT link, resulting in a significant decline in the accuracy of the conventional fixed channel estimation algorithms, and failing to meet the requirements of energy beam optimization and position tuning. To address this issue, channel estimation techniques for MA need to establish the relationship between antenna positions and the corresponding WPT channel responses while maintaining low pilot overhead. In addition, by combining the evolutionary characteristics of space-time channels, dynamic channel prediction can be achieved. It can effectively address channel aging and latency issues caused by antenna movement, and achieve high-precision channel sensing under the constraint of low pilot overhead. Overall, this technology can provide real-time and accurate channel state support for MA position optimization and spatial energy focusing, establishing a complete closed-loop optimization system of dynamic WPT.

\section{Case Study}
\label{sec:5}

To demonstrate the potential of MA in WPNs, we consider an MA-enabled SWIPT system with nonlinear logistic energy harvesting. The objective is to minimize the total transmit power under the constraints of minimum harvested power, minimum data rate and minimum inter antenna distance by jointly optimizing the MA position, transmit beamforming vectors, and power splitting ratio.

    Consider a base station equipped with multiple MA, where each user has separate information decoding and energy harvesting receivers. To focus on how to minimize transmit power, we assume that the perfect channel state information (CSI) can be obtained. Due to the multi-variable coupling and non-convexity in the constraints, it is challenging to directly solve this problem. Therefore, we develop a tractable AO framework that successively optimizes these three variables. In each iteration, the beamforming, power splitting, and antenna position subproblems are solved using appropriate convex approximation techniques, and the process is repeated until convergence. This framework illustrates how antenna positioning can be integrated with conventional SWIPT transmission design to jointly exploit spatial and signal domain resources.

    In the process of numerical simulation, the network is considered with $M$ MAs, whose positions are optimized within a predefined \(\text{1}\,\text{m} \times \text{1}\,\text{m}\) two-dimensional region. There are $K$ users randomly distributed in a square region of size \(\text{1}\,\text{m} \times \text{1}\,\text{m}\), and the center of the user region is \(\text{12}\,\text{m}\) away from the center of the antenna region. A multi-path field-response channel model is adopted. In order to demonstrate the effectiveness of the proposed algorithm, we compare it with the following benchmark scheme: The FPA is a benchmark scheme where antennas are fixed at predetermined locations. The Antenna Selection Matrix (ASM) is a benchmark scheme proposed by antenna selection matrix technology. The Zero-Forcing (ZF) is a benchmark scheme that employs zero-forcing beamforming. 

    Fig. 3 shows the required transmit power versus the minimum rate requirement for the Proposed Algorithm, FPA, and ASM. In Fig. 3, as expected, increasing the minimum rate requirement raises the transmit power of all schemes. Nevertheless, the proposed MA-based joint optimization consistently achieves the lowest transmit power, demonstrating the benefit of exploiting favorable spatial channel conditions through antenna repositioning. Fig. 4 illustrates the influence of varying the minimum harvested-power requirement on required transmit power. It can be seen from Fig. 4 that although all schemes require higher transmit power as the minimum harvested-power requirement increases, the Proposed Algorithm maintains a clear performance advantage. FPA scheme cannot fully exploit the available spatial flexibility.
    These results highlight that MA’s native spatial reconfiguration capability is the core source of performance gains, and jointly optimizing antenna positions and transmission parameters fully unlocks this spatial DoF to improve the performance of MA‑enabled SWIPT systems.
    
\begin{figure}[t]
\centering
\includegraphics[width=0.48\textwidth]{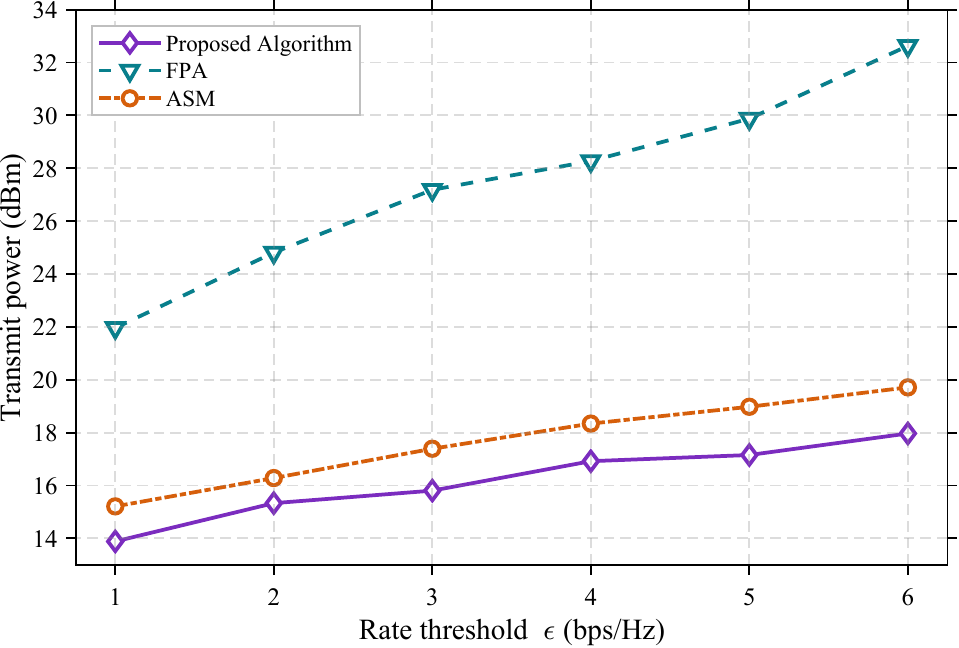} 
\caption{Transmit power versus transmit power budget.}
\label{fig:accuracy}
\end{figure}

\begin{figure}[t]
\centering
\includegraphics[width=0.48\textwidth]{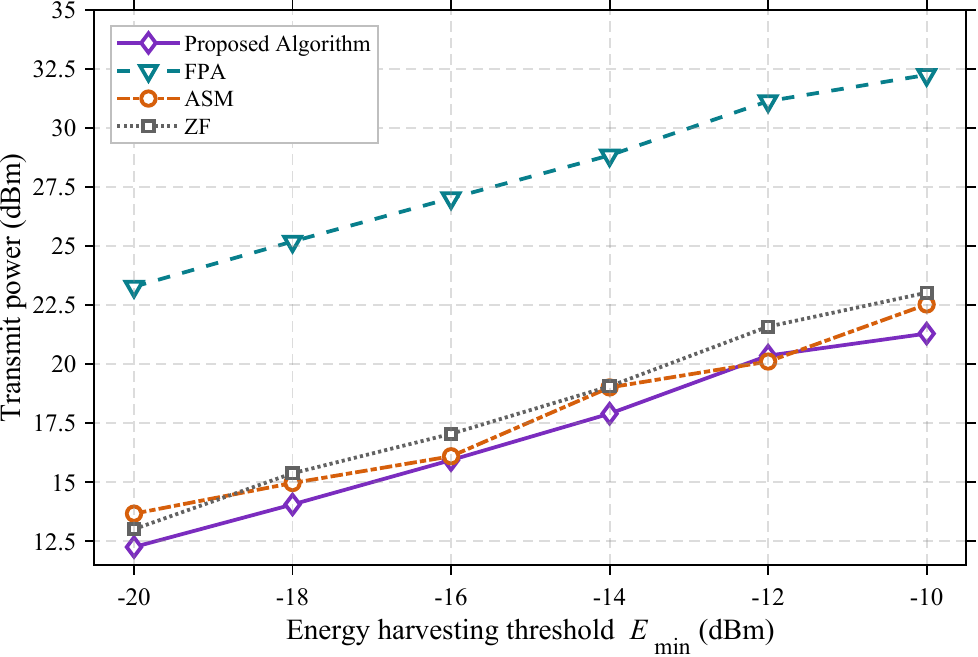}
\caption{Transmit power versus the minimum harvested-power requirement.}
\label{fig:beams}
\end{figure}

\section{Open Challenges and Future Directions}
\label{sec:6}

Although MA-enabled WPNs bring promising performance improvements by exploiting spatial DoF, a series of practical challenges still restrict their real deployment. This section summarizes key open issues and potential research directions in three aspects: adaptive antenna positioning, robust beamforming design, and practical constraints on MA mobility. We analyze the bottlenecks in each dimension and outline corresponding future research directions.

\subsection{Adaptive MA Positioning in WPNs}
By allowing the antenna units to dynamically adjust their physical positions in a preset limited area, MA technology can effectively use the spatial DoF of the wireless channel to improve the communication performance. In WPNs, MA can dynamically adjust its position to optimize the WPT link. However, in dynamic environments, such as scenarios where the receiving devices move rapidly or the channel changes rapidly, achieving low complexity and efficient adaptive antenna positioning is a core challenge. The existing researches mainly focus on the antenna position optimization based on instantaneous CSI. However, this usually leads to time delay and energy consumption caused by real-time antenna movement, which makes it face challenges in practical application. Frequent MA position adjustment will increase the complexity of operation, especially in the fast fading channel, where the acquisition of real-time CSI may face great difficulties. In order to solve the above problems, lightweight algorithms that can balance performance and computational complexity should be explored in the future. In addition, antenna position optimization using long-term or statistical channel information can reduce the dependence on real-time CSI, thus reducing the energy consumption and time delay of frequent movement. Future work may develop predictive MA positioning schemes tailored for WPNs. Instead of reacting to instantaneous channel updates, these schemes can exploit statistical channel knowledge and user mobility traces to pre‑compute antenna positions. This helps reduce unnecessary mechanical movements, cut positioning latency, and improve net harvested energy under practical MA motion-consumption constraints.

\subsection{Robust Beamforming in MA-Enabled WPNs}
Beamforming is one of the key technologies for MA-enabled WPNs to improve WPT efficiency. By accurately controlling the signal direction, the system can concentrate WPT to the target receiving device to minimize energy loss. However, the imperfection of CSI, hardware damage and multi-user interference may lead to a significant decline in beamforming performance, thereby reducing the WPT efficiency. Many works usually assume that CSI is perfect, but in actual systems, there are always errors or delays in the acquisition of CSI, which will lead to serious degradation of the system beamforming performance. The actual antenna array may also have hardware defects such as excessive phase noise, which will affect the accuracy of beamforming to a certain extent. Besides, in multi-user WPNs, simultaneous WPT for multiple receiving devices may cause interference between different beams and affect the efficiency of WPT. To solve the above problems, a robust beamforming algorithm considering CSI uncertainty, hardware damage and interference can be developed in the future to ensure that the system performance can still be guaranteed in the worst case. Furthermore, data-driven deep learning approaches can be investigated for MA‑enabled WPNs, where both antenna positions and beamforming weights are jointly learned from real world channel datasets. This enables robust energy beamforming under CSI imperfections and multi‑user interference, targeting the maximization of net harvested energy rather than generic communication performance.

\subsection{MA Mobility Constraints in WPNs}
MA technology optimizes the wireless channel by flexibly adjusting the antenna positions, but this mobility will introduce additional power loss and time delay for WPNs, which may offset the performance gain it brings. During the operation of the system, the mechanical or electronic scanning of MA array will consume additional energy, thus reducing the overall energy efficiency of the network. The energy consumption of antenna movement needs to be deducted from the total system gain, which further offsets the performance gain brought by mobility. In addition, it takes time for the antenna to move from one position to another, which will bring fixed time delay to WPNs. In high-speed mobile scenarios or applications with high real-time requirements, excessive time delay may lead to outdated CSI, thereby affecting the accuracy of beamforming, and even causing communication interruptions. Therefore, for MA-enabled WPNs, it is necessary to develop a low-power MA drive mechanism and control strategy in the future to reduce power loss by optimizing the movement trajectory, reducing unnecessary movement or using materials and drive technology with higher energy efficiency. Meanwhile, a scheduling algorithm considering the delay of antenna movement can also be designed, such as caching data during antenna movement, to reduce the impact of delay on performance. Through in-depth study on the dynamic trade-off relationship between performance gain, power loss and delay caused by antenna movement, the optimal system designs and operation strategies in different application scenarios of WPNs are hoped to be found in the future.

\section{Conclusion}
\label{sec:7}

This article provided an overview of the framework of MA-enabled WPNs. Compared with the traditional system equipped with FPAs, the introduction of MA brings additional spatial DoF for WPNs. This improves the WPT efficiency, stability and interference resilience of the system, and gives the system more flexible and adaptive beamforming capabilities. In addition, we have verified the effect of MA on reducing the transmit power of WPNs through a specific case study. Based on the above advantages, it can better adapt to the dynamic and complex environment, and has broad application prospects in industrial IoT, smart agriculture, emergency communication and smart home. Since the research on MA-enabled WPNs is still in its infancy, it needs to be further explored and improved in many aspects. It is hoped that this paper can provide valuable reference for future theoretical research and specific practices.

\balance
\bibliographystyle{IEEEtran}
\bibliography{reference}
 
\end{document}